\documentstyle[aps,prb]{revtex}
\makeatletter
\@addtoreset{equation}{section}
\makeatother

\begin{document}
\title{On Lieb's conjecture \thanks{This paper is the
preprint FT-180-1979 which appeared twenty years ago. 
I have submitted it at that time for publication to the journals 
Letters in Mathematical Physics and Communications in Mathematical
Physics. It was refuted by both these
journals and remained unpublished until now. Recently I have
discovered the LANL electronic preprint math-ph/9902017 by Peter Schupp
entitled
"On Lieb's conjecture for the Wehrl entropy of Bloch coherent states"
in which some of the results of my paper are rediscovered.
This fact determined me to emphasize the
existence of my preprint by converting it into an electrronic preprint.}}
\author{Horia Scutaru}
\address{Department of Theoretical Physics, Institute of Atomic Physics,\\
PO Box MG-6,\\ Bucharest-Magurele, Romania,\\
e-mail: scutaru@theor1.theory.nipne.ro}
\maketitle
\begin{abstract}
The reformulation of Lieb's conjecture,
in the frame of the harmonuic analysis on the $SO(3)$
group, makes it evident that the exact value of the
classical entropy of a pure quantum state, which
belongs to the Hilbert space ${\cal H}_{J}$ of a
$(2J+1)$-dimensional, unitary, irreducible representation
${\cal U}_{J}$ of the $SO(3)$ group, depends only on the
parameters which characterize the orbits of ${\cal U}_{J}$
in ${\cal H}_{J}$. In the case $J=1$ we give the exact
analytic dependence of the classical entropy of a quantum state
on the parameters which characterize the orbits and as a
consequence we obtain a verification of Lieb's entropy
conjecture. We verify this conjecture also for any value
of $J$ for the states of the canonical basis of ${\cal H}_{J}$.
A natural generalization of Lieb's entropy conjecture,
which is a new "phenomenon" in the harmonic analysis on
$SO(3)$, is discussed in the case $J=1$.
\end{abstract}
\small
\section{Introduction}

The present paper is devoted to a verification of Lieb's entropy
conjecture \cite{lieb} in some particular cases. 
From the begining we point out the connection of this problem
 with the harmonic analysis on the group $SO(3)$ of rotations
in three dimensions. Let ${\cal U}_{J}(g)$, ($g \in SO(3)$)
be a unitary irreducible representation of $SO(3)$ in the
$(2J+1)$-dimensional Hilbert space ${\cal H}_{J}$, where
$J={1 \over 2}, 1, {3 \over 2},...$, and let $\{v_{m}\}$,
$m=-J, -J+1,..., J-1, J$, be the canonical basis in ${\cal H}_{J}$.
We denote by $||.||$ the norm in ${\cal H}_{J}$ and suppose that
$||v_{m}||=1$ for any value of $m$. The matrix elements of the
representation ${\cal U}_{J}$ in the canonical basis are denoted
by :
\begin{equation}
t_{mn}^J(g)=(v_{m},{\cal U}_{J}(g)v_{n})=\exp\{-i(m\phi+n\psi)\}
t_{mn}^J(\theta)
\end{equation} 
where $(\phi, \theta, \psi)$ are the Euler angles which define the
rotation $g \in SO(3)$, and 
\begin{equation}
t_{mn}^J(\theta)=P_{mn}^J(\theta)
\end{equation}
where
\begin{eqnarray}
&&
\nonumber
P_{mn}^J(\cos\theta)=i^{-m-n}\sqrt{{(J-m)!(J-n)! \over (J+m)!(J+n)!}}
(ctg{\theta \over 2})^{m+n}\\
&&
\nonumber
\sum_{k=max(m,n)}^J {(-1)^k(J+k)! \over (J-k)!(k-m)!(k-n)!}
(\sin{\theta \over 2})^2k\\
\end{eqnarray}
Because the unitary, irreducible representations of the compact
groups are square integrable we have in the particular case of the
$SO(3)$ group, for any $u,v \in {\cal H}_{J}$:
\begin{equation}
{2J+1 \over 8 \pi} \int_{0}^{{2 \pi}} \int_{0}^{{2 \pi}} \int_{0}^{{2 \pi}}
|(u,{\cal U}_{J}(g)v)|^2 \sin \theta d \theta d \phi d \psi=
||u||^2 ||v||^2
\end{equation}
and from this, for any $u \in {\cal H}_{J}$, we obtain :
\begin{equation}
{2J+1 \over 4 \pi} \int_{0}^{{2 \pi}} \int_{0}^{{2 \pi}}
|(u,{\cal U}_{J}(g)v_{\pm J})|^2 \sin \theta d \theta d \phi =
||u||^2 
\end{equation}
Lieb's conjecture takes in these notations the following form :
\begin{equation}\label{en}
-{d \over dp}
({2J+1 \over 4 \pi} \int_{0}^{{2 \pi}} \int_{0}^{{2 \pi}}
|({u \over ||u||},{\cal U}_{J}(g)v_{\pm J})|^2 \sin \theta d \theta d \phi )
|_{p=1} \geq {2J \over 2J+1}
\end{equation}
where the equality is attained only for the Bloch coherent
states :
\begin{equation}\label{bloch}
u={\cal U}_{J}(g)v_{\pm J}
\end{equation}
for any $g \in SO(3)$. In fact this conjecture may be considered
as a consequence of the following conjecture :
\begin{equation}\label{conj}
{2J+1 \over 4 \pi} \int_{0}^{{2 \pi}} \int_{0}^{{2 \pi}}
|(u,{\cal U}_{J}(g)v_{\pm J})|^{2p} \sin \theta d \theta d \phi 
 \leq {2J \over 2pJ+1}||u||^{2p}
\end{equation}
where, when $p \geq 1$, the equality is attained 
only for Bloch coherent states (\ref{bloch}), and when $p=1$ for any
$u \in {cal H}_{J}$. 
This last conjecture is in fact a conjecture about the
sharp estimation of the $L^{2p}(S^2)$-norms of the matrix
coefficients $(u,{\cal U}_{J}(g)v_{\pm J})$:
\begin{equation}
({2J+1 \over 4 \pi} \int_{0}^{{2 \pi}} \int_{0}^{{2 \pi}}
|(u,{\cal U}_{J}(g)v_{\pm J})|^{2p} \sin \theta d \theta d
\phi)^{{1 \over 2p}} 
 \leq ({2J \over 2pJ+1})^{{1 \over 2p}}||u||
\end{equation} 
A result of this kind is unknown in the harmonic analysis on the
$SO(3)$ group. For the Heisenberg group such a result was proved
in \cite{scut1}. In section 2 we obtain the exact value of the
classical entropy of a quantum state \cite{lieb,scut2} and as a
consequence we verify (\ref{en}) for $J=1$.
In section 3 we obtain the exact value of the left hand side of
(\ref{conj}) and prove (\ref{en}) and (\ref{conj}) for any value
of $J$, for the states of the canonical basis: $\{u=v_{m}; m=-J,
-J+1,...,J-1,J\}$. In section 4 we discuss the conjecture (\ref{conj})
for $J=1$.

\section{The exact value of the classical entropy of a quantum
state for J=1}

The essential property of the integral:
\begin{equation}
I_{p}^J(O_{u})=
{2J+1 \over 4 \pi} \int_{0}^{{2 \pi}} \int_{0}^{{2 \pi}}
|(u,{\cal U}_{J}(g)v_{\pm J})|^{2p} \sin \theta d \theta d
\phi,
\end{equation}
which we shall exploit in the following, is the fact that it
depends only on the orbit $O_{u}$ of ${\cal U}_{J}$ in 
${\cal H}_{J}$ to which beelongs the vector $u \in {\cal H}_{J}$.
From this property it follows that it is sufficient to calculate
the integral $I_{p}^J(O_{u})$ only for one representant from each
orbit $O_{u}$; this representant may be choosen to be the most
simple one. The space ${\cal H}_{J=1}$ splits \cite{daum}
into three strata (union of orbits with the same stabilizer up to
conjugacy)  which are characterized by a real valued parameter $a
\in [0,1]$ which is defined for any vector $u=c_{-1}v_{-1}+
c_{0}v_{0}+c_{1}v_{1} \in {\cal H}_{1}$ in the following way
\begin{equation}
a(u)={|c_{0}^2-2c_{-1}c_{1}| \over |c_{-1}|^2+|c_{0}|^2+|c_{1}|^2}
\end{equation}
A typical vector $u$ of a stratum, which is characterized by a
given value of this parameter, is of the following form:
\begin{equation}
u=||u||(\sqrt{1-a}v_{-1}+\sqrt{a}v_{0})
\end{equation}
The stratum for which $a=0$ contains only the two-dimensional
orbit $O_{0}=O_{v_{-1}}=O_{v_{1}}$, which is the orbit of Bloch
coherent states. The stratum with $a \in (0,1)$ is a continuous set
of three-dimensional orbits $O_{a}$, one for each value of 
the parameter $a$. The stratum for which $a=1$ contains only the
two-dimensional orbit $O_{1}=O_{v_{0}}$.

We shall obtain the classical entropy \cite{lieb,scut2}
of a pure quantum state 
${u \over ||u||}=(\sqrt{1-a}v_{-1}+\sqrt{a}v_{0})$, defined by:
\begin{equation}\label{clen}
{\bf S}^{cl}({u \over ||u||})=-{d \over dp}I^{1}_{p}(O_{a})|_{p=1}
\end{equation}
as a function of $a \in [0,1]$. With the notation $x=\cos\theta$
we have:
\begin{equation}
I^{1}_{p}(O_{a})={3 \over 4 \pi}\int_{-1}^{1}dx \int_{0}^{2 \pi}d\phi
\left[(1-a)({1-x \over 2})^2+2a({1-x \over2})({1+x \over 2})+
2(2a(1-a)({1+x \over 2})({1-x \over 2})^3)^{{1 \over 2}}\cos(\phi+
{\pi \over 2})\right]
\end{equation}
For $a=0$ and $a=1$ we obtain:
\begin{equation}\label{s0}
I_{p}^{1}(O_{0})={3 \over 2p+1}
\end{equation}
and
\begin{equation}\label{s1}
I_{p}^{1}(O_{1})={3 \over 2p+1}{2^p \Gamma(p+1)^2 \over \Gamma(2p+1)}
\end{equation} 
respectively.

For each $a \in(0,1)$ we split the integral with respect to $x$
in two pieces: one from $-1$ to ${1-3a \over 1+a}$ and other
from ${1-3a \over 1+a}$ to $1$. Further we change the variable
in the first integral into $t={2a(1+x) \over (1-a)(1-x)}$ and in the
second integral into $t={(1-a)(1-x) \over 2a(1+x)}$. Then after
the integration with respect to $\phi$ and the use of the formula
\cite{bat}:
\begin{equation}
F(-p,-p;1;t)={1 \over 2 \pi}\int_{0}^{2 \pi}(1+2t\cos(\phi + \alpha)+
t^2)^p d \phi
\end{equation}
we obtain
\begin{equation}\label{inte}
I^{1}_{p}(O_{a})={3 \over 2}\left[{(1-a)^{p+1} \over 2a}\int_{0}^{1}
dt({1-a \over 2a}t+1)^{-2(p+1)}F(-p,-p;1;t) +
{(2a)^{2p+1} \over (1-a)^{p+1}}\int_{0}^{1}dt(1+{2a \over 1-a}t)^{-2(p+1)}
t^p F(-p,-p;1;t) \right]
\end{equation}
From the fact that the integrands which appear in (\ref{inte})
are free of singularities for $t \in [0,1]$, $a \in (0,1)$ and
$p \geq 1$, it follows that $I^{1}_{p}(O_{a})$ is a differentiable
function of $a$ and $p$ for $a \in (0,1)$ and $p \geq 1$.
We shall calculate the classical entropy (\ref{clen}) using 
this representation for $I^{1}_{p}(O_{a})$.
From the fact that
\begin{equation}
F(-p,-p;1;t)=1 + p^2t+({p(p-1) \over 2!})^2t^2 + ({p(p-1)(p-2)
\over 3!})^2t^3 + ...
\end{equation}
and because this series is absolutely converging for all 
$t \in [0,1]$ we obtain :
\begin{equation}
{d \over dp}F(-p,-p;1;t)|_{p=1}=2t
\end{equation}
for all $t \in [0,1]$. After tedious calculations we obtain
the following simple expression for the classical entropy of
a pure quantum state ${u \over ||u||} \in O_{a}$ :
\begin{equation}\label{gen}
{\bf S}^{cl}(a)= {2 \over 3}+(a -ln(1+a))
\end{equation}
for $a \in (0,1)$. When $a=0$ or $a=1$ we obtain directly
from (\ref{s0}) and (\ref{s1})
\begin{equation}\label{s00}
{\bf S}^{cl}(0)= {2 \over 3}
\end{equation}
and
\begin{equation}\label{s11}
{\bf S}^{cl}(1)= {2 \over 3}+(1 -ln2)
\end{equation}
respectively.
We remark that (\ref{s00}) and (\ref{s11}) are particular cases
of (\ref{gen}) for $a=0$ and $a=1$ respectively. The relation
(\ref{gen}) is thus valid for all $a \in [0,1]$. Now Lieb's
entropic conjecture for $J=1$:
\begin{equation}
{\bf S}^{cl}(1) \geq {2 \over 3}
\end{equation}
where the equality is attained only for the Bloch coherent
states ${u \over ||u||} \in O_{0}$, is a simple consequence
of (\ref{gen}), (\ref{s00}) and of the well known inequality:
\begin{equation}
a-ln(1+a) \geq 0
\end{equation}
which is valid for all $a \geq 0$. From (\ref{gen}) it is obvious
that the classical entropy attains its maximum value for $a=1$,
i.e., for ${u \over ||u||} \in O_{v_{0}}$.

\section{The verification of Lieb's conjecture and of its
generalization for any value of $J$ for the vectors
of the canonical basis}

We shall calculate the exact value of the integrals 
$I_{p}^{J}(O_{v_{m}})$ for $m=-J, -J+1 ,..., J-1, J$,
where $J={1 \over 2},1,{3 \over 2},2,...$ and $p \geq 1$.
In this case we have
\begin{equation}
I_{p}^{J}(O_{v_{m}})={2J+1 \over 2}\int_{-1}^{1}|P^{J}_{m, -J}(x)
|^{2p} dx={2J+1 \over 2}\int_{-1}^{1}|P^{J}_{m, J}(x)
|^{2p} dx
\end{equation}
and obtain
\begin{equation}\label{exac}
I_{p}^{J}(O_{v_{m}})={2J+1 \over 2pJ+1}({(2J)! \over (J+m)!(J-m)!})^p
{\Gamma(p(J-m)+1)\Gamma(p(J+m)+1) \over \Gamma(2pJ+1)}
\end{equation}
From this fornula it is obvious that $I_{p}^{J}(O_{v_{m}})=
I_{p}^{J}(O_{v_{-m}})$ for all values of $m$. The classical
entropy of a pure quantum state $v_{m}$, $m=-J, -J+1,...,J-1,J$,
is then given by:
\begin{eqnarray}
&&
\nonumber
{\bf S}^{cl}(v_{m})=(J+m)\left({1 \over J+m+1}
+{1 \over J+m+2}+...+{1 \over 2J}
\right)+\\
&&
\nonumber
(J-m)\left({1 \over J-m+1}+{1 \over J-m+2}+...+
{1 \over 2J}\right)-
\ln({(2J)! \over (J+m)!(J-m)!})+{2J \over 2J+1}\\
\end{eqnarray}
Since
\begin{equation}
{\bf S}^{cl}(v_{-J})={\bf S}^{cl}(v_{J})={2J \over 2J+1}
\end{equation}
it follows that Lieb's entropic conjecture is then equivalent
with the following inequality in which we have used the
notations $k=J+m$, $j=J-m$ :
\begin{equation}\label{ineq}
k({1 \over k+1}+{1 \over k+2}...+{1 \over k+j})+j({1 \over j+1}
+{1 \over j+2}+...+{1 \over j+k}) \geq \ln({(k+j)! \over k! j!})
\end{equation}
For $k=1$ we obtain a well known inequality \cite{mitri}
\begin{equation}
{1 \over 1}+{1 \over 2}+{1 \over 3}+...+{1 \over j} \geq \ln(j+1)
\end{equation}
valid for any nonnegative integer $j$. This inequality is proved
by induction, using the well known inequality:
\begin{equation}\label{exm}
{1 \over k} \geq \ln({k+2 \over k+1})
\end{equation}
which is valid for any nonnegative integer $k$. We can also prove
the inequality (\ref{ineq}) by induction, first with respect to $k$
and finally with respect to $j$, using (\ref{exm}). In this way we
have proved Lieb's entropic conjecture for any value of $J$ and for
all states $v_{m}$, $m=-J, -J+1,...., J-1, J$. We remark that
with the use of inequality (\ref{exm}) we may prove that 
${\bf S}^{cl}(v_{m})$ attains it maximum value for $m=0$.

In the following we shall discuss the generalized conjecture:
\begin{equation}
I_{p}^{J}(O_{v_{m}}) \leq {2J+1 \over 2pJ+1}
\end{equation}
for any value of $J$ and for $p \geq 1$.
From (\ref{exac}) it follows that this inequality is
equivalent with the following inequality for the $\Gamma$-function:
\begin{equation}
{\Gamma(kp+1)\Gamma(jp+1) \over \Gamma((k+j)p+1)} \leq
\left({\Gamma(k+1)\Gamma(j+1)\over
\Gamma(k+j+1)}\right)^{p}
\end{equation}
which is unknown. This inequality may be written as an inequality
for the $B$-function:
\begin{equation}
((k+j)p+1)B(kp+1,jp+1) \leq ((k+j+1)B(k+1,j+1))^{p}
\end{equation}
for any nonnegative integers $k$ and $j$ and any $p \geq 1$.
We shall consider the most general inequality:
\begin{equation}
((a+b)p+1)B(ap+1,bp+1) \leq ((a+b+1)B(a+1,b+1))^{p}
\end{equation}
for any real nonnegative numbers $a$ and $b$ and for any $p \geq
1$. We have a proof of this inequality only for $a=b$. This is based
on the integral representation for the $B$-function \cite{krat}
(see \S 1.1,1.6.3) which in the case $a=b$ becomes:
\begin{equation}
{1 \over (2b+1)B(b+1,b+1)}=2^{2b}\int_{-{\pi \over 2}}^
{{\pi \over 2}}(\cos \phi)^{2b} {d\phi \over \pi}
\end{equation}
Then from Jensen's inequality \cite{rudin}
(see chap. 3, Th. 3.3) we have:
\begin{eqnarray}
&&
\nonumber
\left({1 \over (2b+1)B(b+1,b+1)}\right)^{p} =\\
&&
\nonumber
\left(2^{2b}\int_{-{\pi \over 2}}^
{{\pi \over 2}}(\cos \phi)^{2b} {d\phi \over \pi}\right)^{p}
\leq \\
&&
\nonumber
2^{2pb}\int_{-{\pi \over 2}}^
{{\pi \over 2}}(\cos \phi)^{2pb} {d\phi \over \pi}=\\
&&
\nonumber
{1 \over (2bp+1)B(pb+1,pb+1)}\\
\end{eqnarray}
Hence, the inequality for $a \neq b$ remains a conjecture.

\section{Discussion of the generalized conjecture in the case
$J=1$}

In this section we shall discuss, for $J=1$, the conjecture
(\ref{conj}) which in this case becomes :
\begin{equation}\label{conj1}
I_{p}^{1}(O_{a}) \geq {3 \over 2p+1}
\end{equation}
for any value of $a \in (0,1)$ and for any $p \geq 1$.
In order to verify (\ref{conj1}) we try to find the
explicit form of the integral $I_{p}^{1}(O_{a})$ as
a function of the parameter $a$. First we calculate
this integral, in a straightforward manner, in the case
in which $p$ is a positive integer ($p=n \geq 1$), and
obtain :
\begin{equation}
I_{n}^{1}(a)={ 3 \over 2n+1}\sum_{s=0}^{[{n \over 2}]}
\sum_{r=0}^{n-2s}\sum_{t=0}^{n-s-r}{(-1)^t2^{s+r}
(2n-s-r)!(s+r)!n!a^{s+r+t} \over (s!)^2r!(n-s-r)!(2n)!
(n-s-r-t)1t!}
\end{equation}
After tedious calculations we obtain from this expression
that the coefficients of $a^{2k+1}$ are equal to zero for
$k=0,1$ and that the coefficients of $a^{2k}$ for $k=1,2$
are of the following form:
\begin{equation}
I_{n}^{1}(a)={ 3 \over 2n+1}\left(1-{n(n-1) \over 2(2n-1)}a^2+
{n(n-1)(n-2)(n-3) \over 2^2 2(2n-1)(2n-3)}a^4-...\right) 
\end{equation}
The comparison of this expression with the following function:
\begin{equation}
{2^n(n!)^2 \over (2n)!}a^n P_{n}({1 \over a})=\sum_{k=1}^{[{n
\over 2}]}{(-1)^k n(n-1)(n-2)...(n-2k+1) \over
2^k k! (2n-1)(2n-3)...(2n-2k+1)}a^{2k},
\end{equation}
where $P_{n}(\cdot)$ are the Legendre polynomials,
suggests that:
\begin{equation}\label{sug}
I_{n}^{1}(a)={ 3 \over 2n+1}{2^n(n!)^2 \over (2n)!}a^n 
P_{n}({1 \over a}).
\end{equation}
If we assume that (\ref{sug}) is valid we obtain that:
\begin{equation}\label{ineq1}
I_{n}^{1}(a) \leq { 3 \over 2n+1}
\end{equation}
where the equality is attained only for $a=0$.
Indeed from the fact that thee roots of the Legendre
polinomials lie all in the interval $(-1,1)$ and from
the fact that if $P_{n}(b)=0$ it results that either $b=0$
or $P_{n}(-b)=0$, we obtain that:
\begin{equation}
P_{n}(x) \leq {(2n)! \over 2^n (n!)^2}x^n
\end{equation}
for any $x > 1$. From this inequality we get:
\begin{equation}
{2^n(n!)^2 \over (2n)!}a^n 
P_{n}({1 \over a}) \leq 1
\end{equation}
which together with (\ref{sug}) gives (\ref{ineq1}).
Now we shall try to extend the formula (\ref{ineq1}) to
all real values of $p \geq 1$ using spherical functions
$P_{p}(x)$ instead of the Legendre polynomials.
Then we shall make the hypothesis that:
\begin{equation}
I_{p}^1(O_{a})={3 \over 2p+1}{2^p \Gamma(p+1)^2 \over \Gamma(2p+1)}
a^p P_{p}({1 \over a})
\end{equation}
Then, from the fact that \cite{grad} :
\begin{equation}
P_{p}(z)=({1+z \over 2})^pF(-p,-p;1;{z-1 \over z+1})
\end{equation}
for $Re(z) > 0$, we obtain:
\begin{equation}
I_{p}^1(O_{a})={3 \over 2p+1}{2^p \Gamma(p+1)^2 \over \Gamma(2p+1)}
({1+a \over 2})^pF(-p,-p;1;{1-a \over 1+a})
\end{equation}
From this formula we obtain immediately that:
\begin{equation}
{dI_{p}^1(O_{a}) \over dp}|_{p=1}={2 \over 3}+(a-\ln(1+a))
\end{equation}
{\bf which coincides with the result proved in section II.}

Finally we remark that the inequality (\ref{conj1}) is equivalent
with the following inequality:
\begin{equation}
P_{p}(x) \leq {\Gamma(2p+1) \over 2^p \Gamma(p+1)^2} x^p
\end{equation}
for all $x \geq 1$, or with the inequality:
\begin{equation}
F(-p,-p;1;t) \leq {\Gamma(2p+1) \over 2^p \Gamma(p+1)^2}(1+t)^p
\end{equation}
for all $t \in [0,1]$. We do not have a proof for these
two last inequalities, which are unknown, for noninteger values
of $p$.
%\newpage

\end{document}